\documentclass[1p,pdftex]{elsarticle}
\usepackage{subfigure}
\pdfoutput=1

\title{Lemon: an MPI parallel I/O library for data encapsulation using LIME}
\author[a]{Albert Deuzeman}
\address[a]{Albert Einstein Center for Fundamental Physics, Institute for Theoretical Physics, University of Bern, Sidlerstrasse 5, CH-3012 Bern, Switzerland}
\author[b]{Siebren Reker}
\address[b]{Centre for Theoretical Physics, University of Groningen, Nijenborgh 4, 9747 AG, Groningen, the Netherlands}
\author[c]{Carsten Urbach}
\address[c]{Helmholtz-Institut f{\"u}r Strahlen- und Kernphysik (Theorie) and Bethe Center for Theoretical Physics, Universit{\"a}t Bonn, 53115 Bonn, Germany}
\author{for the ETM Collaboration}
\begin{document}

\begin{frontmatter}

\begin{abstract}
We introduce Lemon, an MPI parallel I/O library that is intended to allow for efficient parallel I/O of both binary and metadata on massively parallel architectures. Motivated by the demands of the Lattice Quantum Chromodynamics community, the data is stored in the SciDAC Lattice QCD Interchange Message Encapsulation format. This format allows for storing large blocks of binary data and corresponding metadata in the same file. Even if designed for LQCD needs, this format might be useful for any application with this type of data profile. The design, implementation and application of Lemon are described. We conclude with presenting the excellent scaling properties of Lemon on state of the art high performance computers.


\end{abstract}

\end{frontmatter}
\begin{small}
\noindent
{\bf PROGRAM SUMMARY}\\
{\em Program Title:} Lemon                                    \\
{\em Journal Reference:}                                      \\
{\em Catalogue identifier:}                                   \\
{\em Licensing provisions:} GNU General Public License (GPLv3)\\
{\em Programming language:} MPI and C                         \\
{\em Computer:} Any which supports MPI I/O                    \\
{\em Operating system:} Any                                   \\
{\em Has the code been vectorised or parallelised?:} Yes      \\
{\em RAM:} Depending on input used.                           \\
{\em Number of processors used:} Varies, tested up to 32768   \\
{\em Supplementary material:}                                 \\
{\em Classification:} 11.5                                    \\
{\em External routines/libraries:} MPI                       \\
{\em Nature of problem:} Lattice QCD large file format I/O performance \\
{\em Solution method:} MPI parallel I/O based implementation of LIME format  \\
{\em Restrictions:} \\
{\em Running time:} Varies depending on file and architecture size, in the order of seconds\\
{\em References:}
\end{small}

\section{Introduction}

The tremendous increase in computing power available to the scientific community has brought along with it a large increase in parallelised applications, where many computing cores work together on a single problem. One class of these parallel applications implements a Markov Chain Monte-Carlo (MCMC) algorithm. This algorithm produces a set of data files (known as \emph{configurations}) which serve as the basis for an analysis of quantities one is interested in. That analysis might again be run on a massively parallel system, though not necessarily at the time of generation of the configurations. Usually the data for this algorithm is distributed over all participating compute cores, hence this type of application has to solve the problem of efficiently writing and reading those configurations to and from disk. Moreover, it is important to store metadata, especially problem and algorithm related parameters, unambiguously together with the data itself. This is not only important for avoiding errors in the analysis chain, but also allows for a straightforward distribution of data among researchers, for example by means of grid facilities.

A prominent branch of research implementing an MCMC algorithm is Lattice Quantum Chromodynamics (LQCD), that studies the interactions of quarks and gluons on a four-dimensional discretised spacetime \emph{lattice} as a controlled approximation to the continuum theory. Historically, the data files generated within the LQCD community have been a prised commodity. They represent both a considerable investment of computer resources by the collaborations that generated them and a rich resource for the investigation of QCD. With the availability of increased computer power and the demand for more precise simulations, the size of configurations has steadily grown over time. Since 1997, US collaborations have made their gauge configurations publicly available through the National Energy Research Scientific Computing Center (NERSC) Gauge connection~\cite{NERSC:1997st}. While it was (and still is) simple and effective, it was not the ideal solution for broad international use and contribution. The development of a truly international lattice data grid (ILDG) began several years later at the Lattice conference in 2002, as a result of discussions organised by the UK-QCD collaboration \cite{Davies:2002mu}. The existence of a storage grid generated the need for unification of file formats and descriptions of metadata. Since 2005, the standardised file format (the ILDG file format \cite{ildg:2005st}) describes a specific organisation of binary data and minimal requirements on the provided metadata, encapsulated in SciDAC's Lattice QCD Interchange Message Encapsulation (LIME) file format. Together with the publicly available description of the latter, a reference implementation in C of an interface to such files -- known as C-LIME \cite{lime:2005st} -- was made available by the US-QCD collaboration. This reference implementation has since been incorporated in the majority of LQCD codes. 

The typical size of a data file produced by codes run within the LQCD community has increased from tens of megabytes to several gigabytes presently and there is no indication of this growth slowing down. This increase became possible by the excellent scaling properties of the computational part of modern LQCD codes. A fundamental result on scaling known as Amdahl's law~\cite{amdahl:1967aa} states that the extent to which code scales well is limited by that part of it which is serial. The I/O offered by C-LIME is a serial component, based on regular POSIX\footnote{POSIX is the Portable Operating System Interface [for Unix] IEEE 1003 and ISO/IEC 9945 standard} I/O. The European Twisted Mass Collaboration (ETMC) had been using C-LIME and began to observe an increasing impact of its I/O procedures on the overall scaling as simulations became larger, revealing the urgency of parallelisation of the I/O routines. 

The parallelisation of application programs is often achieved using the Message Passing Interface (MPI) standard. In the second iteration of this standard, usually referred to as MPI-2 and published in 1997, parallelised I/O was introduced. By now this standard is implemented on most high performance computation systems~\cite{top:500}. The current revision of the MPI-2 standard \cite{forum:2009} begins its chapter on I/O with:
\begin{quote}
\noindent \it POSIX provides a model of a widely portable file system, but the portability and optimisation needed for parallel I/O cannot be achieved with the POSIX interface.
\end{quote}
This is the essential reason for the creation of the Lemon library, made available by us and discussed in this work: it provides an interface for the LIME file format based upon the parallel I/O facilities offered by MPI-2. We describe its implementation, discuss some results of performance tests and explain how to integrate Lemon in existing code~\footnote{A very brief introduction to Lemon was already given in \cite{Deuzeman:2009zz}}. 

The development of Lemon addresses demands of the LQCD community and these applications serve both as a motivation for this work and a prototype implementation. But despite of this fact, LIME and Lemon both do have a very general scope. LIME provides a straightforward standard for storing large amounts of generic binary data, with metadata encapsulated in the same file; Lemon is an efficient implementation of this standard for modern massively parallel supercomputers. Both can hence be used in contexts different from LQCD and it is our hope that they may in fact prove useful elsewhere. 

The paper is organised as follows: in the following section we shall describe the set-up and the design of the library. In section~\ref{sec:implementation} we discuss the details of the implementation and the most important interfaces. In sectons~\ref{sec:feasibility} and \ref{sec:test} we discuss benchmark results of Lemon, before concluding. In \ref{sec:api} we give a complete overview of the Lemon API and in \ref{sec:started} we discuss sample programs shipping with the library.

\section{Set-up and Lemon Design}\label{sec:design}

The specific details of the LIME file format are inconsequential to the discussion in this paper, but can be found in ref.~\cite{lime:2005st}. In general a LIME packed file consists of a number of messages, each comprised of one or more records. Usually, at least one record consists of a large amount of binary data. Each record is of a certain type, which can be user-defined. For the example of the ILDG file format, the record containing the binary data is labelled \texttt{ildg-binary-data}. The additional records typically contain a small amount of text providing metadata for the binary data that is encapsulated. Example types are the \texttt{ildg-data-lfn}-record, providing the logical file name under which to find the file on the ILDG grid, and the \texttt{ildg-file-format}-record, providing information on physical parameters and storage precision. The size of the metadata typically comes to a total of about one kilobyte, while the amount of binary data is generally orders of magnitude larger than that. This is one reason for concentrating the parallelisation effort on the writing of the binary data record, a second being that the binary data is typically distributed among the different MPI processes, whereas the metadata is identical for all MPI processes.

A previously mentioned example of binary data that is used in the context of LQCD is a gauge configuration. While we focus in our analysis of parallel I/O on gauge configurations, they are not the only large datafiles in use in LQCD. Two other, closely related, types of datafiles are known as sources and propagators. Both sources and propagators come in many different types, and have therefore not been standardised as much as gauge configurations. They are however commonly written in a LIME format. Typical datasizes range from about a $0.25$ to $10$ times the size of the gauge configuration they belong to. A gauge configuration is defined as a $4$-dimensional spacetime lattice with extent $L_x\times L_y \times L_z \times T$ (where we will always choose $L_x = L_y = L_z = T/2$). A configuration contains $4$ complex valued $3\times 3$ matrices at each of the lattice sites, giving a total of $72$ floating point numbers per site\footnote{It is possible to exploit symmetry properties of the matrices to store the data in a more efficient way than we describe here. This affects the absolute times spent in I/O, but not the scaling and speed of parallel I/O.}. This $4$ dimensional array is stored on disk in a predefined linearised way, for instance for ILDG
\begin{equation}
  \label{eq:linearised}
  p(x,y,z,t) = t L_x L_y L_z + z L_x L_y + y L_x + x\ ,
\end{equation}
where $p$ is the position in the file in units of amount of data per site. During a simulation on a parallel machine, the configuration data is divided over the available compute cores in an ordered way, i.e. every core holds a local chunk of size $l_x \times  l_y \times l_z \times t$ of the global amount. The position of this local chunk in the global data set is determined via integer coordinates $(x_c, y_c, z_c, t_c)$ of the core. Hence, the set of cores is mapped to a Cartesian grid with dimensions $n_x, n_y, n_z, n_t$, and we have e.g. $0\leq x_c < n_x$ and $L_x = n_x  l_x$. Note that the ordering in the file and the again linearised ordering in the memory of the application may differ.

For working with this kind of data parallelism, a particularly suitable supercomputer architecture is IBM's BlueGene line, since these machines are structured in a $3$ dimensional torus, with multiple cores at each site in the torus. This allows for a straightforward mapping of the spacetime lattice onto the supercomputer architecture. Since MPI parallel I/O provides facilities for mapping the data from the architecture of the machine to linearised storage, our intention with Lemon was to take full advantage of these facilities.

The design goals for the Lemon library were the following.
\begin{enumerate}
\item On most high performance platforms, a library implementing the MPI-2 standard is available\footnote{Note that availability does not always mean actual usage by users. MPI parallel I/O puts a much larger strain on filesystems and hardware than serial I/O. We have seen more than once that both our benchmark program and Lemon revealed configuration issues in either hardware or software. We therefore advise to coordinate a first large scale test of parallel I/O with support staff.} and well optimised. This led to our decision to use MPI-2 parallel I/O for Lemon and our principle that calls to the Lemon library should mimic MPI calls: for a given but fixed MPI communicator, Lemon calls should be collective.

\item The POSIX I/O implementation of the LIME interface provided in the C-LIME library is fairly universal. To make Lemon more intuitive to C-LIME users and facilitate the porting of applications library, the functionality and routines implemented in C-LIME should be provided by Lemon wherever possible. The Lemon variation of existing C-LIME routines will, however, have their prefix changed from \texttt{lime} to \texttt{lemon}, to avoid potential namespace clashes. One important difference between otherwise equivalent C-LIME and Lemon calls will be the necessary replacement of a POSIX file pointer by an \texttt{MPI\_File} pointer and the necessity of specifying an MPI communicator. An additional, more subtle difference, is that Lemon uses the integer datatype \texttt{MPI\_Offset} to express displacements and offsets, whereas C-LIME uses \texttt{size\_t} or \texttt{int}.

\item In addition to the C-LIME functionality, Lemon should provide special and dedicated routines for operations that can be optimised for MPI-2 parallel I/O.
\end{enumerate}

We want to stress that code calling Lemon cannot be identical to code calling C-LIME, mainly for the reason that Lemon calls are supposed to be collective. The upshot of this requirement, however, is generally a slight simplification of existing codes.

\section{Implementation} \label{sec:implementation}

Let us start the discussion of the Lemon API with the example of an MPI program writing a simple ASCII message to a LIME file.

\begin{verbatim}
#include <stdio.h>
#include <mpi.h>
#include <lemon.h>
#include <string.h>

int main(int argc, char **argv)
{
  MPI_File fp;
  MPI_Comm cartesian;

  /* Needed for the creation of a Cartesian communicator */
  int mpiSize;
  int distribution[] = {0, 0, 0, 0};
  int periods[] = {1, 1, 1, 1};

  LemonWriter *w;
  LemonRecordHeader *h;

  /* The data we want to write */
  char message[] = "LEMON test message";
  MPI_Offset bytes = strlen(message);

  MPI_Init(&argc, &argv);

  /* Create a 4 dimensional Cartesian node distribution */
  MPI_Comm_size(MPI_COMM_WORLD, &mpiSize);
  MPI_Dims_create(mpiSize, 4 /* #dims */, distribution);
  MPI_Cart_create(MPI_COMM_WORLD, 4 /* #dims */, distribution,
                  periods, 1 /* reorder */, &cartesian);

  /* Open the file using the Cartesian communicator just created
     and initialize a writer with it */
  MPI_File_open(MPI_COMM_WORLD, "canonical.test",
                MPI_MODE_WRONLY | MPI_MODE_CREATE, MPI_INFO_NULL, &fp);
  w = lemonCreateWriter(&fp, MPI_COMM_WORLD);

  /* Create a header, write it out and destroy it again */
  h = lemonCreateHeader(1 /* MB */, 1 /* ME */, "lemon-test-text", bytes);
  lemonWriteRecordHeader(h, w);
  lemonDestroyHeader(h);

  /* Write out the small test string defined earlier as the data block */
  lemonWriteRecordData(message, &bytes, w);
  lemonWriterCloseRecord(w);

  /* Close the writer and release resources */
  lemonDestroyWriter(w);
  MPI_File_close(&fp);

  MPI_Finalize();

  return 0;
}
\end{verbatim}

This example has exactly the same functionality as the first example for a C-LIME program as discussed in ref.~\cite{lime:2005st}. As can be seen, first a variable of type \texttt{MPI\_File} is declared and the opening is performed using \texttt{MPI\_File\_open}. Analogous to the C-LIME API, a Lemon writer must next be created. Each message contains a header, created with \texttt{lemonCreateHeader} and written with \texttt{lemonWriteRecordHeader}. The message is then written with \texttt{lemonWriteRecordData}, before the created structures are removed. Finally the file is closed again using \texttt{MPI\_File\_close}.  

As advertised before, all calls to Lemon library functions are collective. Hence, the above example mimics the corresponding C-LIME example almost exactly, except it is independent of the process rank. Only the prefix \texttt{lime} is replaced with \texttt{lemon} and instead of a POSIX file pointer the corresponding \texttt{MPI\_file} type is used. In addition, \texttt{lemonCreateWriter} requires a reference to the \texttt{MPI\_Communicator}. These differences translate to the creation of a \texttt{lemonReader} indicating the main differences to the C-LIME API as follows: 
\begin{verbatim}
LemonReader* lemonCreateReader(MPI_File *fp, MPI_Comm Cartesian)
LemonWriter* lemonCreateWriter(MPI_File *fp, MPI_Comm Cartesian)
\end{verbatim}
Of course, in the above example the advantage of using Lemon as compared to C-LIME is not yet apparent. While the full Lemon API is described in \ref{sec:api}, we shall turn here to Lemon functionality that extends over C-LIME and leads to the alluded advantages. 

In order to provide efficient I/O for large amounts of distributed binary data, an ordering of that data is necessary first. As discussed in section~\ref{sec:design} around eq.~\ref{eq:linearised} we assume that the set of cores is mapped to a Cartesian grid, each core holding a local part of the global data. Such a mapping is most easily generated using MPI communicators and the corresponding routine \texttt{MPI\_Cart\_create}. It will generate a Cartesian communicator, which needs to be passed to to the \texttt{lemonCreateReader} and/or \texttt{lemonCreateWriter} on construction.

A distributed data set ordered in this sense can be both read and written in a parallel fashion using the following two routines
\begin{verbatim}
int lemonReadLatticeParallel(LemonReader *reader,
                 void *data, MPI_Offset siteSize,
                          int const *latticeDims)
int lemonWriteLatticeParallel(LemonWriter *writer,
                  void *data, MPI_Offset siteSize,
                           int const *latticeDims)
\end{verbatim}
where \texttt{data} is a pointer to the local data to be written to the file, \texttt{siteSize} is the size in bytes of the data structure at a single lattice site in bytes. The amount of data available at \texttt{data} must therefore be equal to $l_x\times l_y \times l_z \times t$ times \texttt{siteSize}. The \texttt{latticeDims} array must contain the dimensions of the problem, \emph{i.e.} $L_x$, $L_y$, $L_z$ and $T$, with the ordering matching that of the Cartesian MPI communicator.

While those two routines provide the most basic functionality, there are others providing more features. Firstly, for the linearisation procedure, there is no \emph{a priori} preferred order of traversing the different dimensions. For reasons of compatibility or performance, one may decide to order the dimensions in the Cartesian communicator different from the ordering that one wants to use when storing data. In such a case the routines
\begin{verbatim}
int lemonReadLatticeParallelMapped(LemonReader *reader,
                       void *data, MPI_Offset siteSize,
            int const *latticeDims, int const *mapping)
int lemonWriteLatticeParallelMapped(LemonWriter *writer,
                        void *data, MPI_Offset siteSize,
             int const *latticeDims, int const *mapping)
\end{verbatim}
can be used to map the data as desired. These routines require the same arguments as their basic version described above, but need an additional array of integers providing permutations of the indices. For example, to write data ordered as XYZT, from fastest to slowest index, in memory to a TXYZ ordering on disk, the array {\tt \{3, 0, 1, 2\}} should be supplied. It is important to note that, since the routines have no information about the data per lattice site other than its size in bytes, any internal reordering of the data on a single lattice site will have to be taken care of separately. The routines described above provide scaling and significantly faster I/O, when compared with serial I/O, as we shall demonstrate in section~\ref{sec:test}. However, there may be applications that are so I/O intensive that a significant fraction of time is spent on file access even when using these parallel I/O routines. In that case, it is actually possible to let I/O overlap with computations, by using what is known as non-blocking MPI operations. These operations are performed by functions that do not wait until their specific task has been completed, but return immediately. Since dedicated hardware is often used for communication and I/O, one may be able to engage in other tasks while waiting for the non-blocking tasks to complete. In an optimal case, this can halve the execution time. Lemon provides the following variants:
\begin{verbatim}
int lemonReadLatticeParallelNonBlocking(LemonReader *reader,
    void *data, MPI_Offset siteSize, int const *latticeDims)
int lemonWriteLatticeParallelNonBlocking(LemonWriter *writer,
     void *data, MPI_Offset siteSize, int const *latticeDims)
\end{verbatim}
While a substantial increase in performance for some applications may be achieved, there are implications for the structure of the code. The main constraint is that the memory buffer (\texttt{*data}) provided to the routines must not be accessed before completion of the MPI routines. To check for completion, the following routines only return upon completion (or failure) of the last I/O call (similar to \texttt{MPI\_Wait}).
\begin{verbatim}
int lemonFinishReading(LemonReader *reader)
int lemonFinishWriting(LemonWriter *writer)
\end{verbatim}
Lemon keeps track of the status of these calls and will automatically wait for previous I/O operations to finish before initiating a new call from the same structure. Finally, a non-blocking version of the non-parallelised I/O routines has been provided for consistency.
\begin{verbatim}
int lemonReaderReadDataNonBlocking(void *dest,
   MPI_Offset const *nbytes, LemonReader *reader)
int lemonWriteRecordDataNonBlocking(void *source,
   MPI_Offset const *nbytes, LemonWriter* writer)
\end{verbatim}
Combined versions of the mapped and non-blocking varieties of the I/O routines are also available. A complete description of the Lemon API is provided in \ref{sec:api}, while additional practical information is given in \ref{sec:started}.

\section{Benchmark setup}\label{sec:feasibility}
To both check the quality of the MPI-2 implementation on different machines and investigate the performance potential of MPI parallel I/O versus serial POSIX I/O, we wrote a dedicated benchmarking program. This program writes and reads the amount of binary data associated with configurations of variable sizes to and from disk. In more detail: in the serial POSIX I/O version of our benchmark we first gather the distributed data to a single MPI process which is then writing the full data set to disk. In MPI parallel I/O version all MPI processis write in parallel using the different routines provided by the MPI-2 standard.

The first step in our feasibility study was to verify the proper working of MPI-2 parallel I/O routines on the supercomputers we had access to. 
Upon completion of this first test, we extended our program to enable a performance comparison of various I/O strategies, with the goal to determine the best reading and writing strategy. The different strategies compared here include a serial reading and writing operation and the different modes of writing provided in the MPI-2 standard. For this test, we used LQCD configurations of with extent $L=24$. The result was that \texttt{MPI\_Write\_at\_all} was the most stable call over all the file systems we had access to, while comparible in speed to other parallel I/O calls. Hence, \texttt{MPI\_Write\_at\_all} was used in the design of Lemon, as discussed more specifically in section \ref{sec:design}. 

Next we tested scaling in more detail using only \texttt{MPI\_Write\_at\_all}. In this test, we used the configuration sizes given in table~\ref{tab:datasize}.
\begin{table}[ht]
  \begin{center}
    \begin{tabular*}{.3\textwidth}{@{\extracolsep{\fill}}cc}
      \hline
      $L$ & size[GB] \\
      \hline
      $24$ & $0.38$ \\
      $32$ & $1.21$ \\
      $48$ & $6.12$ \\
      $64$ & $19.3$ \\
      \hline
    \end{tabular*}
    \caption{Configuration size in GB for configurations of size $L^3\times(2L)$ for different values of $L$. Double precision is assumed for floating point numbers and all elements of the $SU(3)$ matrices are explicitly represented.}
    \label{tab:datasize}
  \end{center}
\end{table}
We performed this scaling test on Jugene~\cite{jugene:2011}, an {IBM} {Blue Gene/P} housed in Forschungszentrum J\"ulich, Germany. This machine is divided into $72$ racks of compute nodes. A single rack is used for development, and $71$ racks are used for production. A rack consists of $1024$ compute nodes with $4$ compute cores each. I/O is performed by dedicated I/O nodes attached to the racks, where the development rack has $32$ I/O nodes attached to it ($1$ I/O node per $32$ compute nodes) and the production racks have $8$ I/O nodes each (one I/O node per $128$ compute nodes). Hence the number of I/O nodes is the relevant dependent quantity in a scaling test. Table~\ref{tab:partsize} shows the number of I/O nodes per possible partition size. Any job requesting a partion size less than half a rack ($512$ nodes) will run on the development rack, a job requesting $512$ nodes or more will run on a production rack. 

Note that these two available number of I/O nodes lead to two possible ways of testing with $4$ and $8$ I/O nodes. Partitions with either $128$ or $512$ compute nodes will each use $4$ I/O nodes. Similarly, partitions with $256$ or $1024$ compute nodes will each have $8$ I/O nodes available to them. This is important for our scaling tests, since going from $256$ to $512$ compute nodes will double the number of compute nodes but halve the number of I/O nodes. We did not test every configuration size on every partition size, because some global lattice sizes do not map well to a large number of nodes and because large lattices on very few nodes will run into physical memory limitations. Table~\ref{tab:partsize} shows which configuration sizes were tested per available partition size.
\begin{table}[ht]
  \begin{center}
    \begin{tabular*}{1.\textwidth}{@{\extracolsep{\fill}}ccccccc}
      \hline
      Compute nodes & Rack type & I/O nodes & $L=24$ & $L=32$ & $L=48$ & $L=64$\\
      \hline
      $32$&dev&$1$& x & x & - & -\\
      $64$&dev&$2$& x & x & - & -\\
      $128$&dev&$4$& x & x & x & -\\
      $256$&dev&$8$& x & x & x & x\\
      $512$ (midplane)&production&$4$& x & x & x & x\\
      $1024$ ($1$ rack)&production&$8$& x & x & x & x\\
      $2048$ ($2$ racks)&production&$16$& - & x & x & x\\
      $4096$ ($4$ racks)&production&$32$& - & x & x & x\\
      $8192$ ($8$ racks)&production&$64$& - & x & x & x\\
      \hline
    \end{tabular*}
    \caption{The number of I/O nodes as function of the partition size for Jugene. A partition of $1024$ nodes is installed as a single rack and therefore this name is often used. Half a rack ($512$ compute nodes) is often called a midplane. The last $4$ columns show which configuration sizes we tested per partition, where "x" indicates that a test was performed, and "-" indicates that a test was not performed. For every "x", between $20$ and $64$ configurations were read and written.}
    \label{tab:partsize}
  \end{center}
\end{table}
The results of this test will be discussed in detail in section~\ref{sec:test}. 

Note that this test uses a stripped down version of Lemon, restricted to the MPI calls involved in the performance critical binary data writing part. To verify that these results accurately represent the performance of Lemon, we have compared the speeds to those obtained using Lemon inside an LQCD code called tmLQCD (described in~\cite{Jansen:2009xp}). The tmLQCD code is a full lattice QCD code employing Lemon for its file I/O, but is also configurable for using C-LIME. Having both implementations allows us to compare the performance of Lemon I/O with both the previous serial implementation, and the stripped down benchmarking program. We have no evidence of any difference in parallel I/O speeds between the dedicated benchmark and the production use of Lemon. 

\section{Benchmark results}\label{sec:test}
The benchmarking code described in section~\ref{sec:feasibility} was run on several different machines, but we report here only on the main scaling test that we performed on the Blue Gene installation in J{\"u}lich. In order to properly analyse scaling, many factors need to be taken into account, including for example network architecture and possible network congestion during I/O processes. Many of these factors are, however, not in our hands and as a consequence the results given here will be system dependent. Nevertheless, the dominant trends are expected to be reproduced elsewhere. We measure performance as a function of the number of I/O nodes. By averaging over many runs we gather statistics and can specify an error, which gives an indication of the effect of the variability in the aforementioned other factors.

\begin{figure}[ht]
  \begin{center}
    \includegraphics[width=\linewidth]{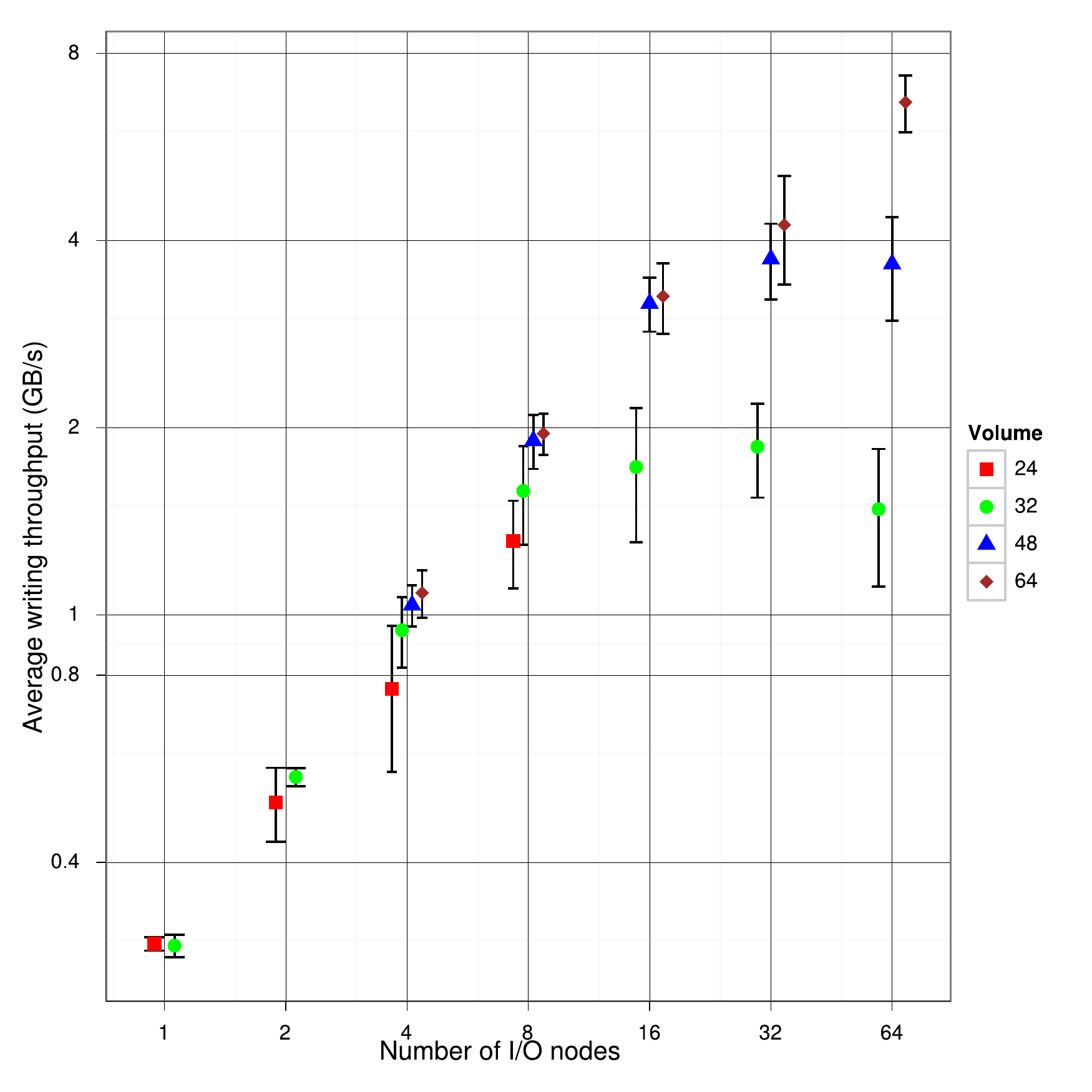}
  \end{center}
  \caption{Average writing speed in $GB/s$ as a function of the number of I/O nodes for the $4$ different lattice sizes given in table~\ref{tab:datasize}. Red squares correspond to $L=24$ configurations, green circles, blue triangles and brown diamonds to $L=32$, $L=48$ and $L=64$ respectively. In order to improve legibility of the plot, points at the same number of I/O nodes have been slightly horizontally displaced. Both the $x$- and the $y$-scale are logarithmic, so perfect scaling would show up as a linear relationship in this plot.}
  \label{fig:jwmean}
\end{figure}

Figure~\ref{fig:jwmean} shows the main results from this scaling test, reporting the average writing speed in GB per second as a function of the number of I/O nodes, for the four different configuration sizes listed in table~\ref{tab:partsize}.

For all tested lattice sizes, we see the same qualitative behaviour: an almost linear increase of the average speed with an increasing number of I/O nodes, saturating from a certain number of I/O nodes on. This saturation can be attributed to the decreasing amounts of data that are available locally. Once this amount becomes too small, the fraction of time spent in the actual writing of this data becomes small as compared to contributions such as network latency or communication overhead. This should not be a problem in practice, since a similar loss of efficiency will be seen in the core computation at these levels of distribution, making them unused in practice.

\begin{figure}[ht]
  \begin{center}
    \includegraphics[width=\linewidth]{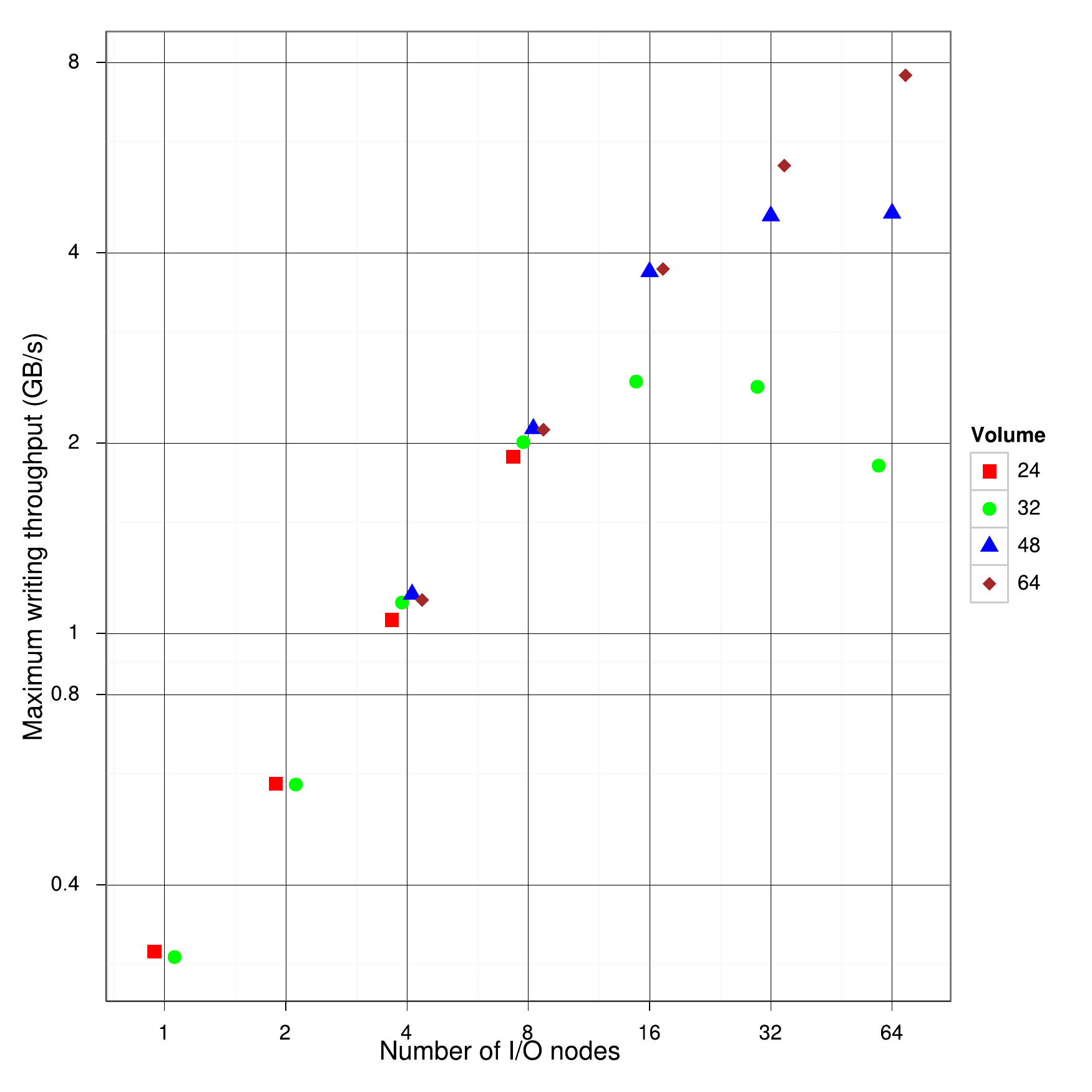}
  \end{center}
  \caption{Maximum writing speed in $GB/s$ as a function of the number of I/O nodes for the $4$ different lattice sizes given in table~\ref{tab:datasize}. Red squares correspond to $L=24$ configurations, green circles, blue triangles and brown diamonds to $L=32$, $L=48$ and $L=64$ respectively. In order to improve legibility of the plot, points at the same number of I/O nodes have been slightly horizontally displaced. Both the $x$- and the $y$-scale are logarithmic, so perfect scaling would show up as a linear relationship in this plot. }
  \label{fig:jwmax}
\end{figure}

In figure~\ref{fig:jwmean} we show average writing speeds. This gives a realistic impression of the use of Lemon in practice because these measurements were sensitive to all external factors that are going to be present in everyday use of the library. A rough estimate of the Lemon performance in a more or less optimal testing condition can be obtained by simply looking at the best achieved I/O times or the fastests writing speeds. This is done in figure~\ref{fig:jwmax}, where we plot only the maximum writing speeds for every setup, of course without error bars. We still see almost linear scaling with the number of I/O nodes until the local volume becomes too small.

In our comparisons with C-LIME, we are forced to choose one master node which directs the writing or offloads the writing to an external dedicated I/O node. We find that the writing speed for this setup is independent of the size of the partition used. This is easily understood, as there is always only one node writing, independent of the partition size. However, one might expect that the size of the written blocks does have an impact on the writing speed. We do not observe such a dependency. In fact, we see writing speeds that are very reproducible, independent of the partition size, block size and even external conditions such as network load. We conclude from this that when using C-LIME in this setup, we are bound by the I/O hardware directly attached to the single writing node. The writing speed we observe is $42.5\pm0.5$ MB/s. This is shown in figure~\ref{fig:jwmean} as the horizontal red line. Parallel I/O is clearly always much faster than the serial I/O, showing the immense potential of the MPI parallel I/O library on machines such as Jugene.

There is one cautionary remark we have to make at this point: on Jugene we observed that the MPI parallel I/O writing calls -- and hence also Lemon writing calls -- do sometimes not write the full set of data. The indication of this to happen is that parts of the created files are filled with only zeros. This problem appears to be not reproducable and we so far did not manage to find its reason. However, since it appears already on the level of MPI routines we are rather certain that it is not a bug in Lemon.

Parallel I/O changes the behaviour of I/O in several respects. We will focus our analysis on the following:
\begin{itemize}
\item Scaling with the partition size (number of I/O nodes)
\item Dependence on the sizes of blocks written
\item Absolute speed increases of up to $100\times$
\item Increased variability of writing speed
\end{itemize}

\subsection{Partition size scaling}
\begin{figure}[ht]
  \centering
  \subfigure[\label{fig:jpartw}]%
  {\includegraphics[width=0.47\linewidth]{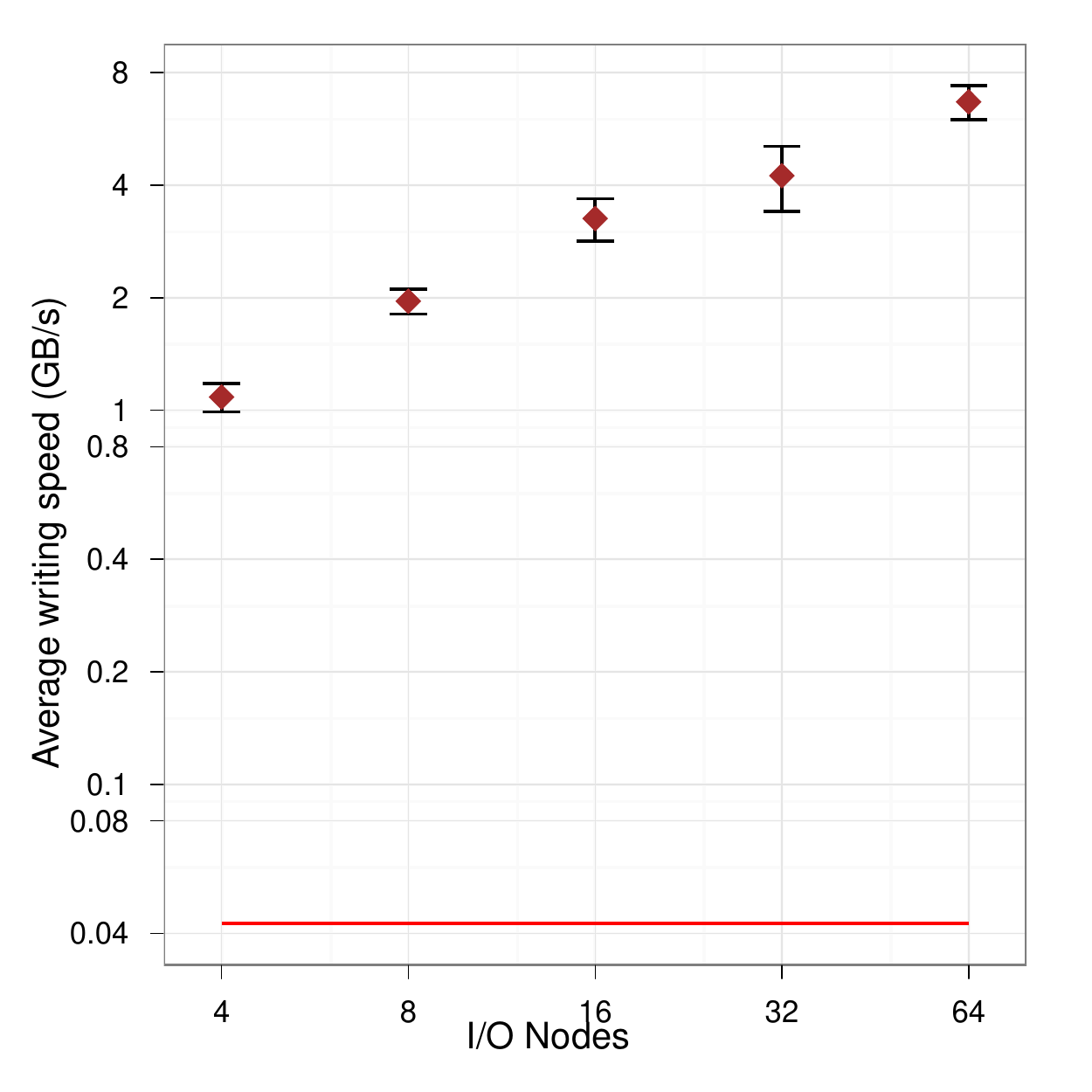}}
  \subfigure[\label{fig:jpartr}]%
  {\includegraphics[width=0.47\linewidth]{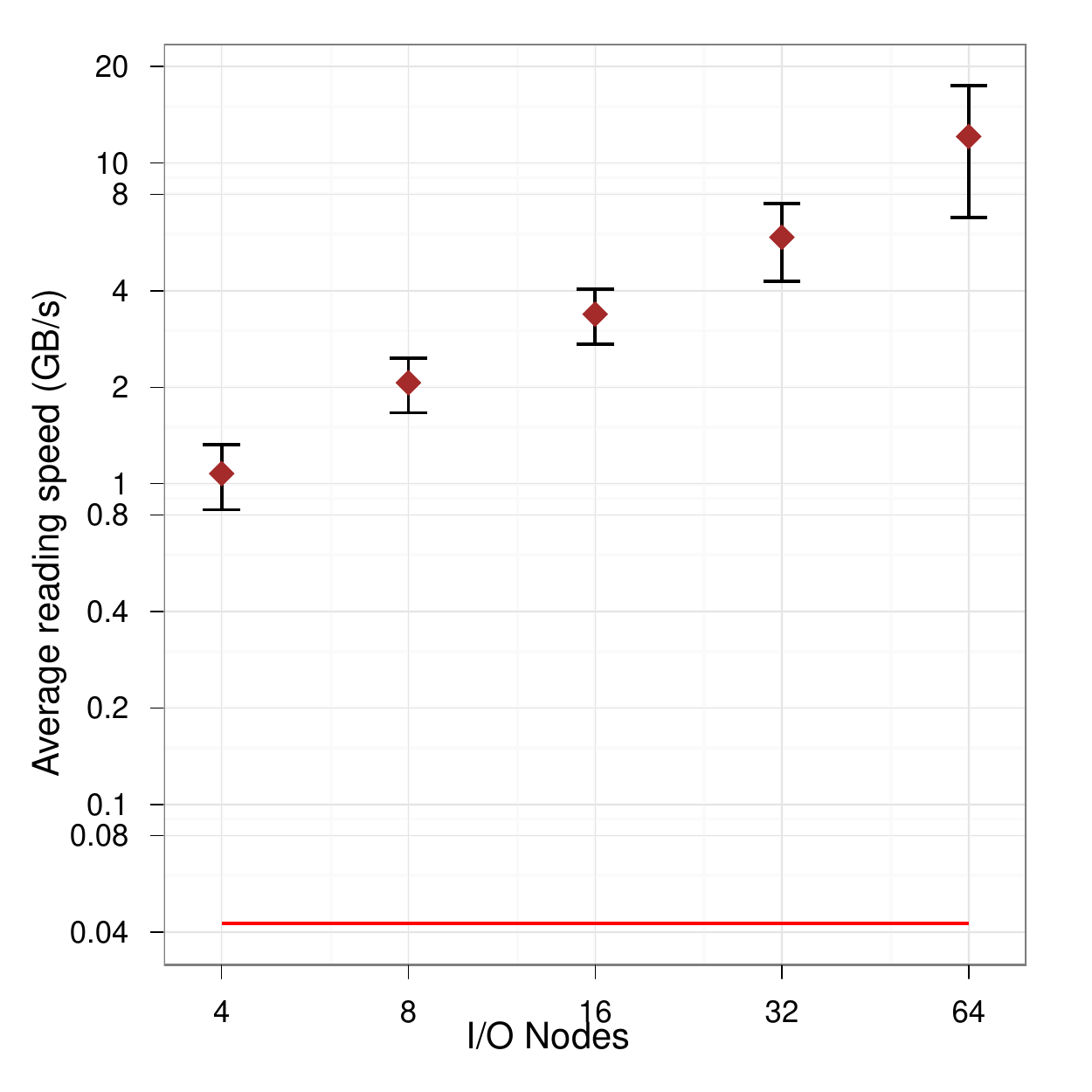}}
  \caption{Average writing (a) and reading (b) speed in gigabytes per second for a $64^3\times128$ configuration, on 4, 8, 16, 32 and 64 I/O nodes. Both the x- and the y-scale are logarithmic, so perfect scaling would show up as a linear relationship in this plot. The red line at the bottom indicates the writing and reading speed for a single I/O node, as measured with C-LIME: $42.5\pm0.5$ megabytes per second.}
  \label{fig:jpart}
\end{figure}

Our main interest in writing Lemon was to obtain parallel I/O with good scaling. We want the I/O to scale with the number of I/O performing nodes, ideally doubling the writing speed when we double the number of I/O nodes. We disregard the influence of block sizes for this comparison, by looking only at the results for a single lattice size (a $64^3\times128$ configuration). Figure~\ref{fig:jpart} shows that we see good scaling for reading (b) and slightly worse scaling for writing (a). This result is reproduced for all lattice sizes, and has also been observed on other systems, though not tested as rigorously elsewhere as it was on Jugene.

\subsection{Block size}
\begin{figure}[!ht]
  \begin{center}
    \includegraphics[width=\linewidth]{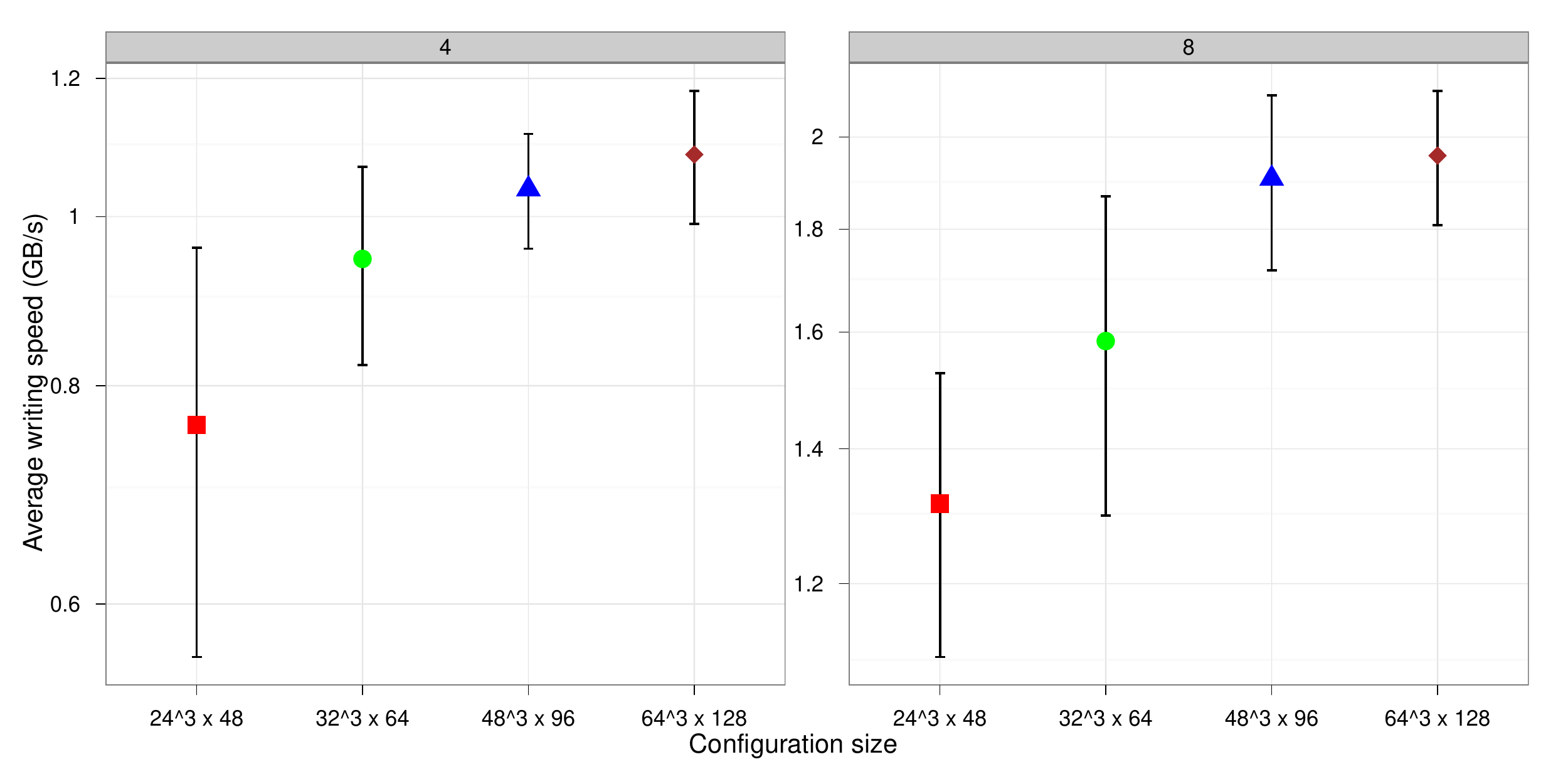}
  \end{center}
  \caption{Average writing speed in gigabytes per second for various configuration sizes while keeping the number of I/O nodes fixed to 4 (left panel) and 8 (right panel).}
\label{fig:jblockw} 
\end{figure}

\begin{figure}[!ht]
  \begin{center}
    \includegraphics[width=\linewidth]{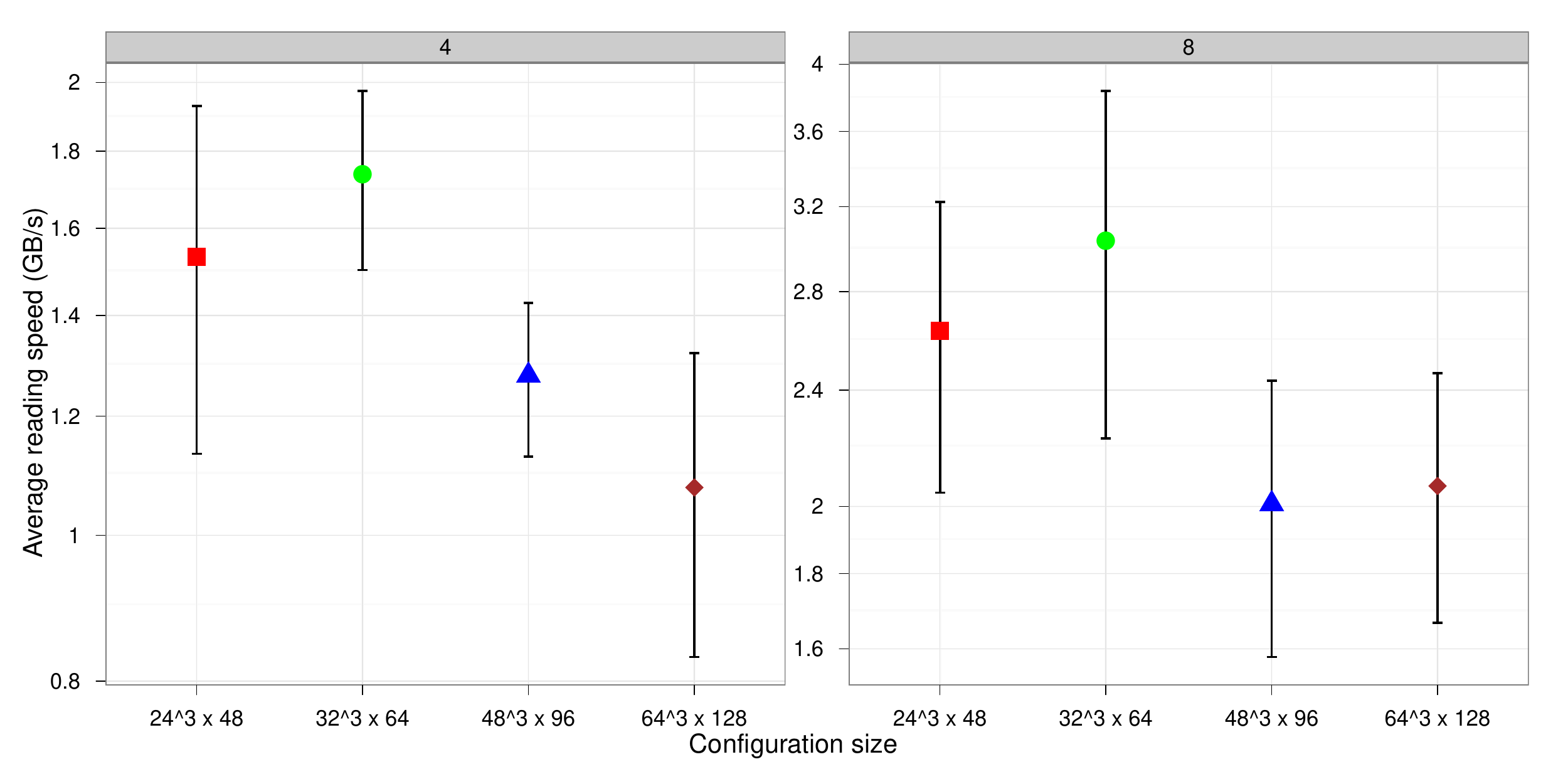}
  \end{center}
  \caption{Average reading speed in gigabytes per second for various configuration sizes while keeping the number of I/O nodes fixed to 4 (left panel) and 8 (right panel).}
\label{fig:jblockr} 
\end{figure}

In the previous section, we wrote the same file size with varying numbers of I/O nodes. Figures \ref{fig:jblockw} and \ref{fig:jblockr} instead show the writing and reading speed for a fixed number of I/O nodes while varying the total file size (and hence the size of the block written by each I/O node). We observe a mild block size dependency for either reading or writing. The writing speed for bigger blocks is a bit faster than the writing speed for smaller blocks. For reading, we observe a different behaviour, where the larger 2 configurations are read at lower speeds than the smaller 2. This is not the behaviour we naively expect, even though the variation is small and the statistical significance is small. However, it is present for both 4 and 8 I/O nodes.

\subsection{Variability}
In our tests we observed a much larger variability of I/O speeds than for the serial case. There are many possible causes for this observation, and most of these are beyond our control in these tests. Examples would be network load by other programs running at the same time as our I/O tests. We actually consider the variability a good thing from a software point of view, as it is an indication that we are maximizing the writing speed, and are constrained by external factors that will also be present in the routine use of Lemon. 

Unfortunately, not every write is faster than in the serial case. We regularly see a slow initial write in our tests. Subsequent writes are never as slow as this initial write, and the problem is not present in the parallel reading of files. From the point of view of Lemon, nothing is changed between the first and any subsequent write or write. We have not observed this behavior on other systems where we have tested, but it has also been observed on this system by other groups\footnote{Private communication with support personnel and other users.}. A separate test of the used MPI subroutines also already revealed slow initial writes. We therefore concluded that Lemon itself is not the source of this behavior, and a solution is beyond the scope of its implementation.

\subsection{Application Performance}

From a practical point of view, we are interested in the actual performance increase comparing C-LIME with Lemon in a typical application and on typical hardware. In this sub-section we present such results, which from our point of view represent the main value of Lemon. The tmLQCD software package~\cite{Jansen:2009xp} is used by the ETM collaboration in production runs and was using C-LIME to read and write configurations to disk. By factorising the code appropriately, we have implemented the Lemon functionality into tmLQCD while maintaining the choice to use C-LIME. The choice is made at configuration time.

There are three typical applications within tmLQCD which we have tested: the first one is the generation of configurations using the so called Hybrid Monte Carlo (HMC) algorithm. Within the HMC I/O appears only relatively rarely, so the expected gain from Lemon is not large. In our test, we measured a single trajectory of the D15.48 run of~\cite{Baron:2010bv} on one rack of Jugene using Lemon and C-LIME. The second application is a matrix inversion using the generated configurations as input and storing the result as output. The result is a vector with $24$ floating point values per lattice site. The gauge configuration had $72$ floating point values per lattice site. In this application I/O can consume a significant fraction of the total run-time, if serial I/O is used. The third application we test is an inversion with multiple shifted matrices (MMS). The computational effort is only about twice that of a single inversion, but the number of results written is typically $10$ to $20$ times larger. Clearly the latter application is where we expect to gain most from using Lemon. For our test we used $18$ shifts and hence had to write $19$ results to disk.

\begin{figure}[!ht]
  \begin{center}
    \includegraphics[width=\linewidth]{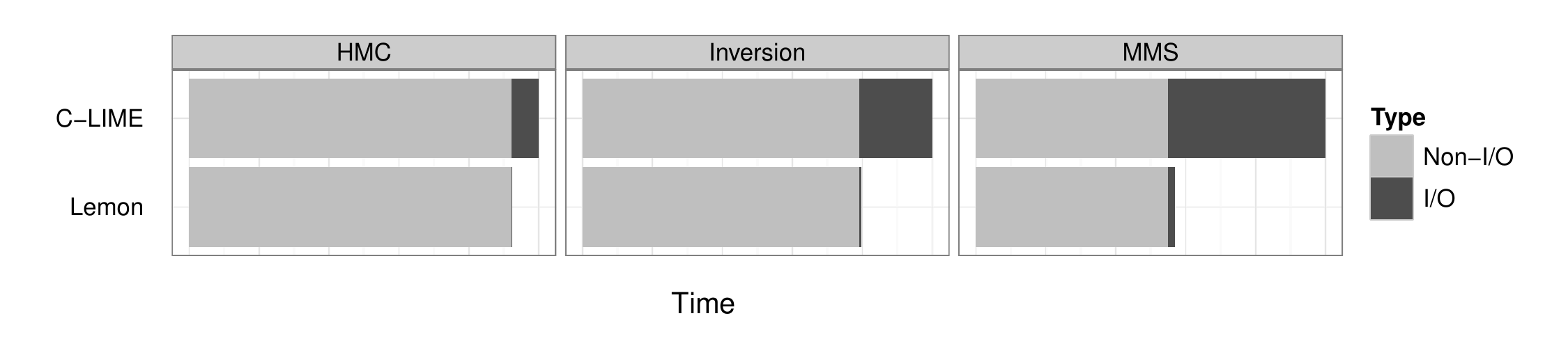}
  \end{center}
  \caption{Timings (in arbitrary units) for representative computations done with the tmLQCD \cite{Jansen:2009xp} package, using either C-LIME or Lemon for the I/O. The three typical computations shown here are Hybrid Monte Carlo (HMC) configuration generation, Inversion (propagator generation) and Multiple Mass Solver inversions.}
\label{fig:tmLQCDimp} 
\end{figure}

The result of our test is visualised in figure~\ref{fig:tmLQCDimp}. We compare the three applications using C-LIME versus Lemon and distinguish the time in arbitrary units spent in file I/O and in other tasks. As expected Lemon reduces the net-time spent in I/O operations significantly compared to C-LIME. While for the HMC application the overall run-time reduction by using Lemon is small, we observe for the inversion application a significant speed-up and for the MMS application Lemon even reduces the run-time to about 57\% of the one with C-LIME. Let us remark that even the relatively small gain with respect to the total run-time observed for the HMC application is rather valuable, since we measure the overall execution time. The net saving in CPU hours is obtained by multiplying by a large number of MPI processes and will be sizeable.

Note that time spent in the computational part of code is strongly dependent on input parameters used in these measurements. It is beyond the scope of this paper to look in more detail at the precise increases for various choices of input parameters. The results shown in figure~\ref{fig:tmLQCDimp} are for typical, but unoptimised parameter values.

\section{Conclusions}
The ILDG LIME format is used in the storage of large lattice QCD gauge configurations. These files are regularly read from and written to disk in lattice QCD computations. An existing library (C-LIME) delivers serial I/O to facilitate this process. We have implemented an MPI parallel I/O library, named Lemon, as a replacement for C-LIME. We have tested the scaling and absolute I/O speeds of Lemon subroutines on Jugene, a BlueGene/P in Forschungszentrum J\"ulich, and performed smaller scale tests on other systems as well. To date, we have tested Lemon on the IBM BG/P at the Rekencentrum of the University of Groningen (the Netherlands), the IBM BG/P Babel at IDRIS in Paris (France), the Huygens POWER 6+ cluster at the SARA center for high performance computing in Amsterdam (The Netherlands), the cluster at HLRN Berlin (Germany), the Jade2 cluster at CINES in Montpellier (France) and the Curie cluster at TGCC in CEA. Lemon is found to scale well, and absolute I/O speed increases of up to $100$ times are regularly observed.

\section{Availability and future development}
Lemon has been made available in the form of a tar-ball from the usual CPC repository. The most recent development snapshot is also available through anonymous SVN access from {\tt svn://thep.housing.rug.nl/lemon}. Some help in compiling and exploring the library can be found in appendix~\ref{sec:started}. Since Lemon, even if it is fundamentally rewritten, is ultimately derived from C-LIME, it inherits its licensing model and is available freely under the conditions of GPLv3 or later. The latest version of the tmLQCD code can be obtained from {\tt http://www.itkp.uni-bonn.de/\~{}urbach/software.html}. While we explicitly allow for an extension of the functionality of Lemon in the future, we see no current need for this. Instead, our plans for the immediate future include primarily an implementation in C++, which should show benefits mainly in the form of a more intuitive interface.

\section*{Acknowledgements}
We wish to thank the support personnel at Forschungszentrum J\"ulich (Germany), the Rekencentrum of the University of Groningen (The Netherlands), IDRIS in Paris (France), the SARA center for high performance computing in Amsterdam (the Netherlands), the cluster at HLRN Berlin (Germany), CINES in Montpellier (France) and TGCC in CEA. The authors wish to acknowledge R\'emi Baron for stimulating discussions, assistance, the access provided to the Babel system in IDRIS and tests performed on the Jade2 cluster. We further wish to thank Karl Jansen for useful comments, and PetaQCD and in particular Gilbert Grosdidier for support, discussions and providing access to the Curie cluster.

\begin{appendix}
\section{The Lemon API}\label{sec:api}

\noindent This appendix gives a complete overview of the Lemon API, describing each function and its parameters.  

\begin{itemize}
\item\verb|LemonReader* lemonCreateReader(MPI_File *fp, MPI_Comm cartesian)|

Creates a \texttt{LemonReader} object, linked to the MPI filepointer \texttt{fp} and using the Cartesian geometry defined by the communicator \texttt{cartesian}. This same communicator should be used in opening the MPI\_File. Memory will be allocated automatically for the object and will have to be freed to avoid memory leaks. In case of allocation failure, the returned value will be \texttt{NULL}.

\item\verb|void lemonDestroyReader(LemonReader *reader)|

Closes the \texttt{LemonReader} object pointed to by \texttt{reader} and frees the memory allocated to it. This function will \emph{not} invalidate the MPI filepointer and Cartesian communicator associated with the \texttt{LemonReader} object on initialisation.

\item\verb|int lemonSetReaderPointer(LemonReader *reader, MPI_Offset offset)|

This particular function, provided for C-LIME compatibility, is a wrapper around \texttt{lemonReaderSeek} and will move the filepointer to the indicated position within the current record. The variable \texttt{offset} indicates a position as an absolute number of bytes, from the start of the record. The return value is an error code, as described at the end of this appendix.

\item\verb|MPI_Offset lemonGetReaderPointer(LemonReader *reader)|

Returns the current position of the filepointer associated with \texttt{reader}, as the number of bytes relative to the start of the current record.

\item\verb|int lemonReaderNextRecord(LemonReader *reader)|

Forwards the filepointer to the start of the next record within \texttt{reader} and parses its header information. This function can only be called when the current record has been properly closed, using the \texttt{lemonReaderCloseRecord} described below, or if this will be the first record in the file. The return value is an error code, as described at the end of this appendix.

\item\verb|int lemonReaderMBFlag(LemonReader *reader)|

LIME uses two specific bits, dubbed the Message Begin (MB) and Message End (ME) bits, to indicate the bounds of a message. This function returns the value of the MB bit for the current record in \texttt{reader}, signalling the start of a new message. By construction, the first record in a LIME file should have the MB bit set to 1. A subsequent non-zero value of the MB bit should only happen after the occurrence of a non-zero value of the ME bit, indicating the end of a message. It is possible for a single record to have both the MB and ME bit set, indicating a message consisting of a single record. Please note that ensuring the consistency of these bits when writing is ultimately the responsibility of the user writing the file.

\item\verb|int lemonReaderMEFlag(LemonReader *reader)|

LIME uses two specific bits, dubbed the Message Begin (MB) and Message End (ME) bits, to indicate the bounds of a message. This function returns the value of the ME bit for the current record in \texttt{reader}, indicating that it is the last record in the message. By construction, the last record in a LIME file should have the ME bit set to 1. Any additional non-zero values of the ME bit should only happen after the occurrence of a non-zero value of the MB bit, indicating the start of a new message. A record with a non-zero MB bit should follow each record with a non-zero ME bit, unless it is the last record in the file It is possible for a single record to have both the MB and ME bit set, indicating a message consisting of a single record. Please note that ensuring the consistency of these bits when writing is ultimately the responsibility of the user writing the file.

\item\verb|char const *lemonReaderType(LemonReader *reader)|

Returns a string to the type of the current record in \texttt{reader}, as specified in the header.

\item\verb|MPI_Offset lemonReaderBytes(LemonReader *reader)|

Returns the size of the data contained in the current record in \texttt{reader}.

\item\verb|MPI_Offset lemonReaderPadBytes(LemonReader *reader)|

Returns the amount of padding added to the current record in \texttt{reader}. Note that padding is handled automatically by Lemon and this function should not be of great importance to the general user.

\item\verb|int lemonReaderReadData(void *dest, MPI_Offset *nbytes,|\\
     \verb|                                   LemonReader *reader)|

Reads up to \texttt{nbytes} bytes of the data in the current record of \texttt{reader}, placing this data at \texttt{dest}. Note that the memory at \texttt{dest} is assumed to have been allocated already. The variable \texttt{nbytes} is modified to give the number of bytes that have been actually read in the operation. This provides the equivalent of a serial read followed by an \texttt{MPI\_Scatter}, meaning that all nodes will have access a copy of the data. For performing reads of distributed data, use the specialised \texttt{lemonReadLatticeParallel} and friends. The return value is an error code, as described at the end of this appendix.

\item\verb|int lemonReaderCloseRecord(LemonReader *reader)|

Closes the current record of \texttt{reader}, waiting for any pending operations to finish if necessary. If this function is not called before \texttt{lemonReaderNextRecord}, the latter function will do so first. The return value is an error code, as described at the end of this appendix.

\item\verb|int lemonReaderSeek(LemonReader *reader, MPI_Offset offset,|\\
     \verb|                                                int whence)|

This function has overlap with \texttt{lemonSetReaderPointer}, in that it will move the filepointer by \texttt{offset}, relative to the point indicated by \texttt{whence}. Valid values for \texttt{whence} are \texttt{MPI\_SEEK\_CUR} (relative to the current position), \texttt{MPI\_SEEK\_SET} (relative to the beginning of the record) and \texttt{MPI\_SEEK\_END} (relative to the end of the record). The return value is an error code, as described at the end of this appendix. An attempt to move the file pointer outside of the current record will result in an error.

\item\verb|int lemonReaderSetState(LemonReader *rdest, LemonReader const *rsrc)|

While this function should have little utility for most users, it is provided for compatibility with the c-lime interface. It copies the state of the reader \texttt{rsrc} to the reader \texttt{rdest}. Note that both readers need to be initialised independently with identical communicators and MPI file handles to the same file.

\item\verb|int lemonEOM(LemonReader *reader)|

Returns 1 if the current record of \texttt{reader} finishes a message, returns 0 otherwise. Note that this is deduced from the MB and ME bits in the header, which in practice cannot always be depended upon to be consistent.

\item\verb|int lemonReadLatticeParallel(LemonReader *reader, void *data,|\\
     \verb|                 MPI_Offset siteSize, int const *latticeDims)|

Reads binary data from the current record in \texttt{reader} associated with a Cartesian grid in parallel mode. At each node, the appropriate chunk of the total data is placed at \texttt{data}. The variable \texttt{siteSize} should contain the size in bytes of the data at a single site of the lattice, while \texttt{latticeDims} is an array containing the global dimensions of the lattice. The local lattice size is calculated automatically, using the Cartesian grid provided to the reader at construction. The return value is an error code, as described at the end of this appendix.

\item\verb|int lemonReadLatticeParallelMapped(LemonReader *reader, void *data,|\\
     \verb|   MPI_Offset siteSize, int const *latticeDims, int const *mapping)|

Equivalent to \texttt{lemonReadLatticeParallel}, but allows for changing the ordering of dimensions between the file and the data in memory by providing an array of permutation indices \texttt{mapping}. This array should be equal in length to \texttt{latticeDims}. The ordering of the dimensions in memory -- \emph{i.e.}, as specified by the Cartesian communicator -- is the one used in specifying both \texttt{latticeDims} and \texttt{mapping}. The indices in the latter should therefore map \emph{from} memory ordering \emph{to} disk ordering.

\item\verb|int lemonReadLatticeParallelNonBlocking(LemonReader *reader,|\\
     \verb|   void *data, MPI_Offset siteSize, int * const latticeDims)|

Equivalent to \texttt{lemonReadLatticeParallel}, but performs the reading in a non-blocking fashion. This means the function will return immediately and allow for performing further calculations in the background. Note that \texttt{data} will be in active use and must therefore not be used until the I/O operation has finished. Checking for the status of this operation is done by calling the \texttt{lemonFinishReading} function, described later. If \texttt{reader} is used in a subsequent Lemon call, the library will also check for completion of the outstanding I/O request automatically and wait for it to finish if necessary.

\item\verb|int lemonReadLatticeParallelNonBlockingMapped(LemonReader *reader,|\\
     \verb|          void *data, MPI_Offset siteSize, int const *latticeDims,|\\
     \verb|                                               int const *mapping)|

Combines the features of \texttt{lemonReadLatticeParallelMapped} and\\ \texttt{lemonReadLatticeParallelNonBlocking}.

\item\verb|int lemonReaderReadDataNonBlocking(void *dest,|\\
     \verb|    MPI_Offset const *nbytes, LemonReader *reader)|

Provides a non-blocking version of \texttt{lemonReaderReadData}. Note that \texttt{data} will be in active use and should therefore not be used until the I/O operation has finished. Checking for the status of this operation is done by calling the \texttt{lemonFinishReading} function, described later. If \texttt{reader} is used in a subsequent Lemon call, the library will also check for completion of the outstanding I/O request automatically and wait for it to finish, if necessary.

\item\verb|int lemonFinishReading(LemonReader *reader)|

If there are any outstanding I/O requests associated with \texttt{reader}, this function will not return until they have been completed. If no such requests are outstanding, or if they have completed already, \texttt{lemonFinishReading} will return immediately. This provides a mechanism for synchronizing calculations and I/O when non-blocking calls are used to overlap both. The return value is an error code, as described at the end of this appendix.

\item\verb|LemonWriter* lemonCreateWriter(MPI_File *fp, MPI_Comm cartesian)|

Creates a \texttt{LemonWriter} object, linked to the MPI filepointer \texttt{fp} and using the Cartesian geometry defined by the communicator \texttt{cartesian}. This same communicator should be used in opening the MPI\_File. Memory will be allocated automatically for the object and will have to be freed to avoid memory leaks. In case of allocation failure, the returned value will be \texttt{NULL}.

\item\verb|int lemonDestroyWriter(LemonWriter *writer)|

Closes the \texttt{LemonWriter} object pointed to by \texttt{writer} and frees the memory allocated to it. This function will \emph{not} invalidate the MPI filepointer and Cartesian communicator associated with the \texttt{LemonWriter} object on initialisation. The return value is an error code, as described at the end of this appendix.

\item\verb|int lemonWriteRecordHeader(LemonRecordHeader const *props,|\\
     \verb|                                      LemonWriter* writer)|

Writes the \texttt{LemonRecordHeader} object at \texttt{props} to initialize a new record in \texttt{writer}. For this to be a valid operation, this should be either the first operation on \texttt{writer}, or it should follow a call to the \texttt{lemonWriterCloseRecord} function described below. The struct at \texttt{props} should be created using \texttt{lemonCreateHeader} (see below) and will then be initialised upon construction. The return value is an error code, as described at the end of this appendix.

\item\verb|int lemonWriteRecordData(void *source, MPI_Offset *nbytes,|\\
     \verb|                                      LemonWriter* writer)|

Writes up to \texttt{nbytes} worth of data from \texttt{source} to the current record of \texttt{writer}. This function is intended for small metadata records and, while being a collective call, performs a serial write from the node with rank 0 in the Cartesian communicator of \texttt{writer}. The return value is an error code, as described at the end of this appendix.

\item\verb|int lemonWriterCloseRecord(LemonWriter *writer)|

Closes the current record in \texttt{writer}, adding padding bytes if required, and frees up the writer for writing the next \texttt{LemonRecordHeader}. The return value is an error code, as described at the end of this appendix.

\item\verb|int lemonWriterSeek(LemonWriter *writer, MPI_Offset offset,|\\
     \verb|                                                int whence)|

Moves the MPI filepointer associated with \texttt{writer} by \texttt{offset} bytes, with \texttt{whence} indicating the position relative to which the offset is taken. Valid values for \texttt{whence} are \texttt{MPI\_SEEK\_CUR} (relative to the current position), \texttt{MPI\_SEEK\_SET} (relative to the beginning of the current record) and \texttt{MPI\_SEEK\_END} (relative to the end of the current record). The return value is an error code, as described at the end of this appendix. An attempt to move the file pointer outside of the current record will result in an error.

\item\verb|int lemonWriterSetState(LemonWriter *wdest, LemonWriter const *wsrc)|

While this function should have little utility for most users, it is provided for compatibility with the c-lime interface. It copies the state of the writer \texttt{wsrc} to the writer \texttt{wdest}. Note that both writers need to be initialised independently with identical communicators and MPI file handles to the same file.

\item\verb|int lemonWriteLatticeParallel(LemonWriter *writer, void *data,|\\
     \verb|                  MPI_Offset siteSize, int const *latticeDims)|

Writes binary data from \texttt{data} to the current record in \texttt{writer} associated with a Cartesian grid in parallel mode. At each node, \texttt{data} should contain the locally available part of the total data. The variable \texttt{siteSize} should contain the size in bytes of the data at a single site of the lattice, while \texttt{latticeDims} is an array containing the global dimensions of the lattice. The local lattice size is calculated automatically, using the Cartesian grid provided to the writer at construction. The return value is an error code, as described at the end of this appendix.

\item\verb|int lemonWriteLatticeParallelMapped(LemonWriter *writer, void *data,|\\
     \verb|    MPI_Offset siteSize, int const *latticeDims, int const *mapping)|

Equivalent to \texttt{lemonWriteLatticeParallel}, but allows for changing the ordering of dimensions between the data in memory and the data in file by providing an array of permutation indices \texttt{mapping}. This array should be equal in length to \texttt{latticeDims}. The ordering of the dimensions in memory -- \emph{i.e.}, as specified by the Cartesian communicator -- is the one used in specifying both \texttt{latticeDims} and \texttt{mapping}. The indices in the latter should therefore map \emph{from} memory ordering \emph{to} disk ordering.

\item\verb|int lemonWriteLatticeParallelNonBlocking(LemonWriter *writer,|\\
     \verb|     void *data, MPI_Offset siteSize, int const *latticeDims)|

Equivalent to \texttt{lemonWriteLatticeParallel}, but performs the writing in a non-blocking fashion. This means the function will return immediately and allow for performing further calculations in the background. Note that \texttt{data} will be in active use and must therefore not be used until the I/O operation has finished. Checking for the status of this operation is done by calling the \texttt{lemonFinishWriting} function, described later. If \texttt{writer} is used in a subsequent Lemon call, the library will also check for completion of the outstanding I/O request automatically and wait for it to finish, if necessary.

\item\verb|int lemonWriteLatticeParallelNonBlockingMapped(LemonWriter *writer,|\\
     \verb|           void *data, MPI_Offset siteSize, int const *latticeDims,|\\
     \verb|                                                int const *mapping)|

Combines the features of \texttt{lemonWriteLatticeParallelMapped} and\\ \texttt{lemonWriteLatticeParallelNonBlocking}.

\item\verb|int lemonWriteRecordDataNonBlocking(void *source,|\\
     \verb|   MPI_Offset const *nbytes, LemonWriter* writer)|

Provides a non-blocking version of \texttt{lemonWriteRecordData}. Note that \texttt{data} will be in active use and must therefore not be used until the I/O operation has finished. Checking for the status of this operation is done by calling the \texttt{lemonFinishWriting} function, described later. If \texttt{writer} is used in a subsequent Lemon call, the library will also check for completion of the outstanding I/O request automatically and wait for it to finish, if necessary.

\item\verb|int lemonFinishWriting(LemonWriter *writer)|

If there are any outstanding I/O requests associated with \texttt{writer}, this function will not return until they have been completed. If no such requests are outstanding, or if they have completed already, \texttt{lemonFinishWriting} will return immediately. This provides a mechanism for synchronizing calculations and I/O when non-blocking calls are used to overlap both. The return value is an error code, as described at the end of this appendix.

\item\verb|LemonRecordHeader *lemonCreateHeader(int MB_flag, int ME_flag,|\\
     \verb|                          char const *type, MPI_Offset reclen)|

Creates a \texttt{LemonRecordHeader} object for use with a writer object. The header is initialised with the values of \texttt{MB\_flag} and \texttt{ME\_flag}, indicating the beginning and ending of a message, the datatype field provided in \texttt{type} and the length of the data in the record in bytes \texttt{reclen}. The value in \texttt{reclen} should only account for the actual data, excluding the header and any padding that will be taken care of automatically. Memory will be allocated automatically for the object and will have to be freed to avoid memory leaks. In case of allocation failure, the return value will be \texttt{NULL}.

\item\verb|void lemonDestroyHeader(LemonRecordHeader *h)|

Frees the memory associated with the \texttt{LemonRecordHeader} object.

\end{itemize}

\noindent The occurence of errors is signalled by the returning of an error code by most functions in Lemon. These error codes are identical to those returned by C-LIME. Checks should be made for a non-zero value of the error code where applicable.

\vspace{0.5cm}
\indent\verb|LEMON_SUCCESS                 =   0|\\
\indent\verb|LEMON_ERR_LAST_NOT_WRITTEN    = - 1|\\
\indent\verb|LEMON_ERR_PARAM               = - 2|\\
\indent\verb|LEMON_ERR_HEADER_NEXT         = - 3|\\
\indent\verb|LEMON_LAST_REC_WRITTEN        = - 4|\\
\indent\verb|LEMON_ERR_WRITE               = - 5|\\
\indent\verb|LEMON_EOR                     = - 6|\\
\indent\verb|LEMON_EOF                     = - 7|\\
\indent\verb|LEMON_ERR_READ                = - 8|\\
\indent\verb|LEMON_ERR_SEEK                = - 9|\\
\indent\verb|LEMON_ERR_MBME                = -10|\\
\indent\verb|LEMON_ERR_CLOSE               = -11|\\

\section{Getting started}\label{sec:started}

This appendix is intended to help new users get Lemon up and running quickly. Upon downloading and extracting the Lemon tarball, one should have a obtained a directory containing, amongst others, a configure script and several subdirectories. These subdirectories include all the headers associated with the library (\texttt{include}), the source code itself (\texttt{src}) and some binaries that do not themselves form part of Lemon (\texttt{check}).

The GNU build system is used for Lemon and the usual \texttt{configure} script is provided. Configuring Lemon should be straightforward, as only the common arguments are taken into consideration. Of relevance is mainly the \texttt{--prefix} argument, that will set the install directory. It is crucial, however, to either use an MPI wrapper around the compiler, or set the include and linker paths such that MPI can be found. This can be done by setting environment variables,  \emph{e.g.} \texttt{CC=mpicc} 

Once the library has been configured, \texttt{make} will compile both the library itself and two binaries that can be found in the \texttt{test} subdirectories\footnote{The system will call \texttt{aclocal-1.9}, \texttt{automake-1.9} and \texttt{autoconf}. This will result in an error if only newer versions are available, but one can simply invoke \texttt{aclocal}, \texttt{automake} and \texttt{autoconf} manually and call \texttt{make} again to continue.}. These two binaries not only provide short, self-contained samples of Lemon usage, but are in fact potentially useful in a production environment. The first, \texttt{lemon\_contents}, is a direct port of the C-LIME \texttt{lime\_contents} program. It displays a short overview of all the records available within a particular LIME file. If the contents of a record data block are both non-binary and short enough, \texttt{lemon\_contents} will send it to standard output. Potential uses for \texttt{lemon\_contents} include checking if a LIME file is well-formed and displaying the metadata associated with a particular file. The second program that is compiled by default is \texttt{lemon\_benchmark}. This executable will generate an artificial lattice of random data, the topology of which will be generated automatically using the number of MPI processes as an input. Its local volume is fixed, so its global volume will scale with the number of processes. This lattice is written out and subsequently read in, both using Lemon's parallel I/O routines. The I/O speed is calculated from the timing of both operations and reported on standard output. To check for the correct operation of the library, an MD5 hash is calculated\footnote{The MD5 implementation included here was written by L. Peter Deutsch and is available freely from \texttt{http://sourceforge.net/projects/libmd5-rfc/files/}.} for the data before and after the I/O operations and any discrepancies will be reported. This code is intended to allow the user to obtain basic information on the I/O performance on his or her particular system and to give some indication of the scaling. Of course, it also functions as a basic tool for detecting major problems in the writing or retrieving of data. After running \texttt{make}, the user can run \texttt{make install} to install the library, its headers and the two binaries described above to the directory specified by the \texttt{--prefix} argument to configure (or the default location if that argument has been omitted). 

Additional binaries can still be compiled by calling \texttt{make check}. These will include the writing and reading of an artifical metadata record, both using blocking and non-blocking I/O (\texttt{xlf} and \texttt{xlf\_non\_blocking}), that can be displayed using \texttt{lemon\_contents}. Also provided are two programs that write a small amount of data using the mapped parallel I/O routines, again in a blocking and non-blocking version (\texttt{parallel} and \texttt{parallel\_non\_blocking}). The data written will be a set of characters identifying a particular MPI process, such that the output file demonstrates the linearisation of the data. The last sample program that has been included is \texttt{canonical}, the example included in section~\ref{sec:implementation}. None of these binaries will be installed, as they serve no particular function other than being examples.

\end{appendix}

\bibliographystyle{plain}
\bibliography{lemon}
\end{document}